\documentclass{aa}
\usepackage{epsf}

\begin{document}
\title{Nucleon superfluidity vs observations of cooling neutron stars}
\author{A. D. Kaminker\inst{1}
\and
P. Haensel\inst{2}
\and
D. G. Yakovlev\inst{1}
}
\institute{Ioffe Physical Technical Institute,
         Politekhnicheskaya 26, 194021 St.~Petersburg, Russia
\and
N.\ Copernicus Astronomical Center,
         Bartycka 18, 00-716 Warsaw, Poland\\
{\em kam@astro.ioffe.rssi.ru, haensel@camk.edu.pl,
 yak@astro.ioffe.rssi.ru}}
\offprints{A.D. Kaminker}

\date{Received x xxx 2001 / Accepted x xxx 2001}
\abstract{
Cooling simulations of neutron stars (NSs) are performed
assuming that stellar cores consist of neutrons, protons
and electrons and using realistic density profiles
of superfluid critical temperatures
$T_{\rm cn}(\rho)$ and $T_{\rm cp}(\rho)$ of neutrons and protons.
Taking a suitable profile of $T_{\rm cp}(\rho)$ 
with maximum $\sim 5 \times 10^9$ K one can
obtain smooth transition from slow to rapid cooling
with increasing stellar mass. Adopting
the same profile one can explain the majority of observations of thermal 
emission from isolated middle--aged
NSs by cooling
of NSs with different masses  either
with no neutron superfluidity in the cores
or with a weak superfluidity, $T_{\rm cn} < 10^8$ K.
The required masses range from $\sim 
1.2 \, {\rm M}_\odot$ for
(young and hot) RX J0822--43 and 
(old and warm) PSR 1055--52 
and RX J1856-3754
to $\approx 1.45\, {\rm M}_\odot$
for the (rather cold) Geminga and Vela pulsars.
Observations constrain the $T_{\rm cn}(\rho)$
and $T_{\rm cp}(\rho)$ profiles with respect to
the threshold density of direct Urca process
and maximum central density of NSs. 
}
\titlerunning{Nucleon superfluidity versus masses in neutron stars}
\authorrunning{A.D. Kaminker et al.}
\maketitle

\section{Introduction}
\label{sect-intro}
In recent years, great progress has been made
in observations of thermal radiation from middle--aged
isolated NSs. Main observational results
are summarized in Fig.\ \ref{fig-cool} which shows
error bars of the effective surface temperature,
$T_{\rm s}^\infty$, as measured by a distant observer,
versus stellar age $t$ for eight sources.
The youngest three are radio-quiet NSs
in supernova remnants, the oldest is also a radio-quiet NS,
while the others manifest themselves  
as radio pulsars. NS ages are mainly pulsar spindown ages
or the estimated supernova ages. 
The age of the Vela pulsar is taken according to Lyne
et al.\ (\cite{lyneetal96}), and the age of
RX J185635--3756 ($\lg t \; [{\rm yr}]=5.95$)
is taken according to Walter (\cite{walter01}).
The values of $T_{\rm s}^\infty$
come from the following sources: 
RX J0822--43 --- Zavlin et al.\ (\cite{ztp99}),
1E 1207--52 --- Zavlin et al.\ (\cite{zpt98}),
RX J0002+62 --- Zavlin \& Pavlov (\cite{zp99}),
PSR 0833--45 (Vela) --- Pavlov et al.\ (\cite{pavlovetal01}),
PSR 0656+14 --- Possenti et al.\ (\cite{pmc96}),
PSR 0630+178 (Geminga) --- Halpern \& Wang (\cite{hw97}), 
PSR 1055--52 --- \"Ogelman (\cite{ogelman95}), and
RX J1856--3754 --- Walter (\cite{walter01}).
For all the sources but the Vela pulsar and 
RX J1856--3754 these data
are presented in Table 3 in Yakovlev et al.\ (\cite{yls99}).
For the Vela pulsar $T_{\rm s}^\infty = 0.68 \pm 0.03$ MK at $1\sigma$ 
confidence level. For RX J1856--3754
we take $T_{\rm s}^\infty = 0.57 \pm 0.06$ MK 
(the central value is mentioned by Walter; we
supplement it with an estimated
error for a joint fit of the optical plus X-ray data).  
The values of $T_{\rm s}^\infty$ 
for the four youngest sources are obtained
from the observed X-ray spectra using hydrogen atmosphere
models. They are more consistent with other information
on these sources (distances, hydrogen column densities,
inferred neutron star radii, etc) than the blackbody
model of NS radiation. On the contrary,
the blackbody model is more consistent for older
sources, and we present the blackbody 
values of $T_{\rm s}^\infty$
for the older sources. 
We do not present observational
data for some other sources. For instance, we do not consider
a  possible NS candidate in Cas A 
(e.g., Pavlov et al.\ \cite{pavlovetal00})
because its nature is not clear.

We will interpret the observational data with the models
of cooling NS. 
NSs of age $t \la 10^6$ yr are known to
cool mainly via neutrino emission
from their interiors, while the older NSs
cool mainly via photon emission from the surface.
We use our fully relativistic nonisothermal
cooling code constructed by O.\ Gnedin and described
in Gnedin et al.\ (\cite{gyp01}). 
We employ the simplest
NS models assuming that the NS cores consist of neutrons (n),
protons (p) and electrons.
We adopt a moderately
stiff 
equation of state (EOS) of this matter
proposed by Prakash et al.\ (\cite{pal88}) (their model I
with the compression modulus of 
saturated
nuclear matter 
 $K=240$ MeV). Other physics input
is the same as described 
in Gnedin et al.\ (\cite{gyp01}).
In particular, we 
place the
core-crust interface at $\rho=1.5 \times 10^{14}$ g cm$^{-3}$.
The outer heat blanketing NS envelope is assumed to be made
of iron (although the atmosphere may contain
light elements)
neglecting the effects of surface magnetic fields;
the relation between $T_{\rm s}$ and
the internal NS temperature is taken from
Potekhin et al.\ (\cite{pcy97}). 
The validity of such approach
is discussed, e.g., by Yakovlev et al.\ (\cite{yls99}). 
The maximum mass of our NS models is $M_{\rm max} = 
1.977\, {\rm M}_\odot$ (with the radius of 10.754 km)
and the maximum central density is $\rho_{\rm c, max}=
2.575 \times 10^{15}$ g cm$^{-3}$.
For the given EOS, powerful direct Urca process
of neutrino emission (Lattimer et al.\ \cite{lpph91})
is forbidden at
$M < M_{\rm D}=1.358 \, {\rm M}_\odot$ ($\rho_{\rm c} < \rho_{\rm D} =
7.851 \times 10^{14}$ g cm$^{-3}$). If so, a NS 
undergoes a {\it slow} cooling
mainly via modified Urca process of neutrino
emission. 
A NS with $M>M_{\rm D}$ possesses a central kernel,
where direct Urca process 
is open ($\rho > \rho_{\rm D}$);  
its cooling is {\it fast}.

The values of $T_{\rm s}^\infty$ for 1E 1207--52,
RX J0002+62, and PSR 0656+14
may be consistent with slow cooling
of non-superfluid NSs. RX J0822--43, PSR 1055-52 
and RX J1856--3754 are too hot
for slow cooling of a non-superfluid NS
with the heat--blanketing envelope made of iron (although
RX J0822--43 is consistent with the blanketing
envelope made of light elements, Potekhin et al.\ \cite{pcy97}).
The lower values of $T_{\rm s}^\infty$, especially
for the Vela and Geminga pulsars, 
are intermediate between slow and fast cooling of non-superfluid
NSs. 
In non-superfluid NSs, the transition
from slow to fast cooling with increasing $M$
occurs in a very narrow mass range $1 < M/M_{\rm D} \la 1.003$
(e.g., Page \& Applegate \cite{pa92}, Yakovlev et al.\ \cite{ykgh01}).
In principle, one can explain the Geminga and Vela data by cooling
of non-superfluid NSs but the probability
that NS masses fall in this very narrow mass range
is too low to accept such explanation. 

We will show that
the observational data in Fig.\ \ref{fig-cool}   
can be explained by the models of NSs
with superfluid cores. 

\section{Nucleon superfluidity}
\label{sect-sup}

We adopt traditional assumption that neutrons
undergo singlet-state Cooper
pairing in the NS crust and triplet-state pairing
in the core, while protons undergo
singlet-state pairing in the core.
Microscopic calculations of the critical
temperatures of the neutron and proton superfluidities (SFs),
$T_{\rm cn}$ and $T_{\rm cp}$, depend sensitively on
the model of nucleon-nucleon interaction and 
many-body theory employed (see, e.g., Yakovlev et al.\ \cite{yls99},
for references). 
All models predict pronounced density
dependence of $T_{\rm cn}$ and $T_{\rm cp}$. In particular,
$T_{\rm cn}(\rho)$ for the singlet--state
neutron SF has maximum at
subnuclear densities in the crust and vanishes
at $\rho \sim 2 \times 10^{14}$ g cm$^{-3}$ while
$T_{\rm cn}(\rho)$ for the triplet--state neutron SF
grows up at subnuclear density, reaches maximum
at $\rho=(2-3) \, \rho_0$
($\rho_0=2.8 \times 10^{14}$ g cm$^{-3}$
is the saturation density of nuclear matter) and decreases
with $\rho$,  vanishing at $\rho \sim (3-5) \times 10^{15}$ g cm$^{-3}$.
$T_{\rm cp}$ also has
maximum at several $\rho_0$ and vanishes at higher $\rho$.
At $\rho \sim (2-3)\, \rho_0$ the proton SF is typically stronger
than the neutron one. SF of n
and p reduces the emissivity of neutrino
reactions and affects heat capacity of n and p.
Moreover, SF initiates a specific 
neutrino emission associated with Cooper pairing of nucleons
(Flowers et al.\ \cite{frs76}).
In this way nucleon SF becomes a strong regulator
of NS cooling.
The appropriate physics input is described in Yakovlev et al.\ (\cite{ykgh01}).

SF of neutrons in the crust affects
cooling of young NSs ($t \la 100$ yr), in which
the internal thermal relaxation is not over,
and it does not affect cooling of middle--aged NSs
we are interested in. 
To be specific, we 
neglect neutron SF in the crust.
As for $T_{\rm cn}(\rho)$ and $T_{\rm cp}(\rho)$ in the
NS core, we will not rely on any particular
microscopic model but will try to find the
models consistent with the observations.
We parameterize the dependence of $T_{\rm c N}$
on the nucleon (N=n or p) wavenumber
$k=k_{\rm F \!N}=(3 \pi^2 n_{\rm N})^{1/3}$ (measured in
fm$^{-1}$) as 
\begin{equation}
    T_{\rm c N}= T_0 \, {(k-k_0)^2 \over (k-k_0)^2 + k_1^2} 
     \; {(k-k_2)^2 \over  (k-k_2)^2 + k_3^2 }~,
\label{Tc}
\end{equation}
for $k_0< k < k_2$, and $T_{\rm c N}=0$ for $k \leq k_0$ and $k \geq k_2$,
$T_0$, $k_0$,\ldots $k_3$ being free parameters.
$T_0$ regulates maximum of $T_{\rm c N}(\rho)$, $k_0$ and $k_2$ determine
low-density and high-density $T_{\rm cN}(\rho)$ cutoffs,
while $k_1$ and $k_3$ govern the shape of $T_{\rm c N}(\rho)$.
Let us present also our simple fit of 
the density dependence of the nucleon number density
$n_{\rm N}$ (in fm$^{-3}$) for the given EOS:
\begin{equation}
    n_{\rm N}=a \, \rho_{14}^b / (1+c \, \rho_{14}+d \, \rho_{14}^2),
\label{nN}
\end{equation}
where $\rho_{14}\equiv \rho /(10^{14}\,{\rm g~cm}^{-3})$;
$a=0.1675$, $b=1.8185$, $c=2.0288$, $d=0.02444$ for N=n, and
$a=0.0006823$, $b=2.6767$, $c=0.1946$, $d=0.01604$ for N=p.

At the first step we neglect neutron SF
in the core and adopt 
$T_{\rm cp}(\rho)$ displayed in Fig.\ \ref{fig-sup}.
It is given by Eqs.\ (\ref{Tc}) and (\ref{nN}) with
$T_0=2.029 \times 10^{10}$ K,  $k_0=0$,
$k_1=1.117$ fm$^{-1}$, $k_2=1.241$ fm$^{-1}$,  $k_3=0.1473$ fm$^{-1}$.
The maximum 
$T_{\rm cp}=6.78 \times 10^9$ K takes place
at $\rho=5.74 \times 10^{14}$ g cm$^{-3}$ and $T_{\rm cp}$ vanishes
at $\rho = 9.55 \times 10^{14}$ g cm$^{-3}$.
This proton SF is constructed artificially but 
is typical for SFs provided
by microscopic calculations. In a not too massive
NS ($M \la 1.45 \, {\rm M}_\odot$, $T_{\rm cp}(\rho_{\rm c}) \ga 10^9$ K) 
it will appear at the initial
stage of NS cooling ($t \la 100$ yr). Accordingly, the peak of
neutrino emission due to Cooper pairing of p
will take place at this early stage making
almost no affect on the cooling curves (solid lines).
Moreover, the internal temperature of the middle-aged NS will be
much lower than $T_{\rm cp}$. The proton SF will
switch off the proton heat capacity and it will strongly suppress
neutrino reactions involving protons,
particularly, modified and direct Urca processes.
However, the total heat capacity is mainly provided by
neutrons being almost unaffected by proton SF.

The upper cooling curve is for the
$1.3 \, {\rm M}_\odot$ NS
(with the radius of 13.04 km) but it would be exactly the same
for all $M$ from $M_{\rm D}$
down to $\sim 1.1 \, {\rm M}_\odot$ (similar insensitivity
to $M$ is discussed, e.g., in Yakovlev et al.\ \cite{yls99}).
For $t \sim (10^2-10^5)$ yr 
this cooling curve goes noticeably higher
than in a non-superfluid NS since SF suppresses neutrino emission and
slows cooling of such a NS.
This allows us to explain
high surface temperature of RX J0822-43 by cooling
of a superfluid NS with the heat blanketing envelope
made of iron.
Similar to the nonsuperfluid NS models, we have 
faster cooling at $M > M_{\rm D}$.
However, the SF suppression
of neutrino emission strongly smoothes transition
from slow to fast cooling
and allows us to explain
the low temperatures $T_{\rm s}^\infty$ of the Vela and Geminga pulsars
by cooling of
NSs with $M \approx 1.47 \, {\rm M}_\odot$. With further increase
of $M$ to $M_{\rm max}$ we obtain almost the same fast-cooling curves
as in the absence of SF.
These results are not sensitive to the specific choice
of $T_{\rm cp}(\rho)$ near the maximum (at $\rho \la \rho_{\rm D}$)  
as long as $T_{\rm cp}(\rho)$
is high (say, $\ga 3 \times 10^9$ K) for $\rho \la \rho_{\rm D}$.
However, the transition from slow to fast cooling with
increasing $M$ is very sensitive to the decreasing slope of
$T_{\rm cn}(\rho)$ at higher $\rho$. By shifting this slope
to higher $\rho$ we would obtain higher masses of the Vela and
Geminga pulsars.

\begin{figure}
\centering
\epsfxsize=86mm
\epsffile[15 145 570 700]{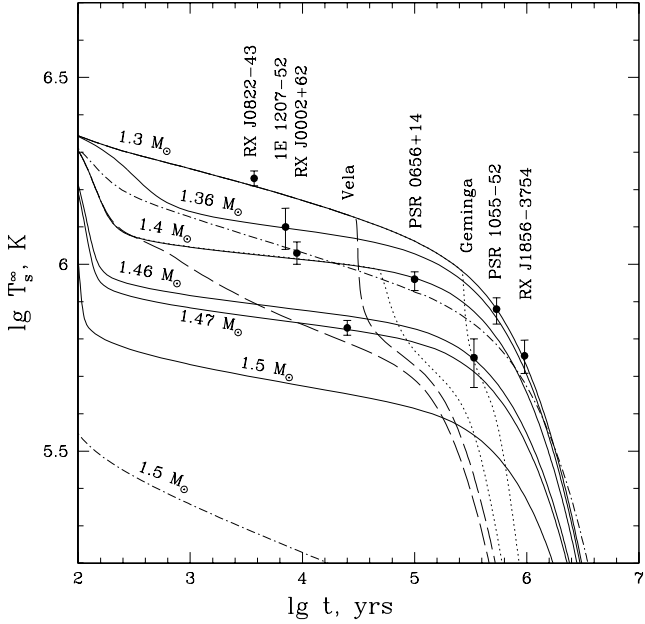}
\caption{
Observational data (error bars) on surface temperatures of eight
NSs (see the text) as compared with theoretical cooling curves
obtained for proton and neutron SFs 
from Fig.\ \protect{\ref{fig-sup}}.
All cooling curves 
(except dot--and-dashed ones)
are calculated for the same proton SF.
Solid lines -- no neutron SF for NS 
models with masses (from top to bottom)
1.3, 1.36, 1.4, 1.46, 1.47, and 1.5 ${\rm M}_\odot$.  
Dashed lines and dotted lines correspond to neutron SFs
for $M=1.3$ and $1.4 \, {\rm M}_\odot$. Dashed-and-dot lines
are for non-superfluid $1.3 \,{\rm M}_\odot$ and $1.5 \, 
{\rm M}_\odot$ NSs.
}
\label{fig-cool}
\end{figure}

\begin{figure}
\centering
\epsfxsize=70mm
\epsffile[15 145 570 700]{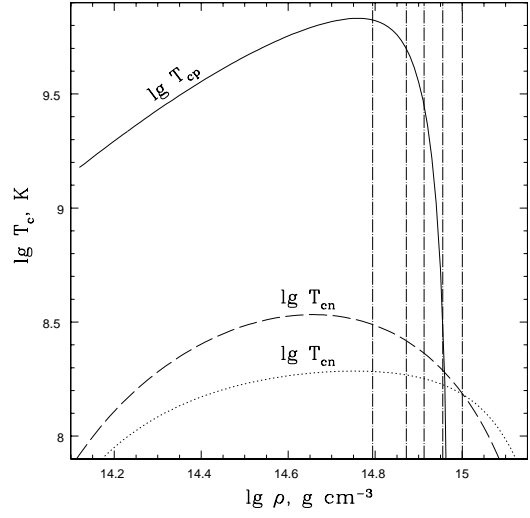}
\caption{
Density dependence for one model of the proton
critical temperature (solid line) and two models of neutron
critical temperatures (dots and dashes) used in cooling
simulations (Fig.\ \protect{\ref{fig-cool}}). Vertical
dot-and-dashed lines show the central densities
of 1.1, 1.3, 1.4, 1.5 and 1.6$\, {\rm M}_\odot$ NSs.
}
\label{fig-sup}
\end{figure}

Our next step is to discuss the effects of neutron SF.
Following the results of microscopic calculations we assume
that this SF is generally weaker than 
the proton one in the NS core. If $T_{\rm cn} \la 3.5 \times 10^8$ K,
this SF appears in the NS core at $t \ga 100$ yr
(for $M \la 1.45\, {\rm M}_\odot$).
The NS core is then isothermal,
and the temperature decreases uniformly over the core
in the course of cooling. Thus
the SF appears first in a spherical layer
at those values of
$\rho=\rho_{\rm CP}$ which correspond to maximum of 
$T_{\rm cn}(\rho)$,
$T_{\rm cn}(\rho_{\rm CP})=T_{\rm cn}^{\rm max}$.
Before that time the cooling curves are
exactly the same (solid lines in Fig.\ \ref{fig-cool})
as in the absence of neutron SF.
Once the neutron SF is switched on,
powerful neutrino emission due to
Cooper pairing of n starts to operate.
If $\rho_{\rm CP}$ is noticeably lower than
the NS central density one can calculate the neutrino
luminosity $L_{\rm CP}$ due to this Cooper pairing
semianalytically. Its temperature dependence
can be written as
$
    L_{\rm CP}(T) = L_0 \,l(\tau),
$
where $L_0$ is a normalization constant (which depends on
$T_{\rm cn}(\rho)$ and NS model) while $l(\tau)$
is a universal function of
$\tau= T/T_{\rm cn}^{\rm max}$ (with $l \propto (1-\tau)^{3/2}$
at $(1-\tau) \ll 1$; $l \propto \tau^8$ at $\tau \ll 1$,
and with maximum at $\tau \approx 0.795$). 
This neutrino emission
cools the star very quickly producing
sharp breaks of the cooling curves.
The breaks are especially pronounced if the
neutron SF appears at $t \sim 10^4-10^6$ yr. 
Soon after the break onsets the cooling curves become even steeper  
due to reduction of the neutron
heat capacity at $T \ll T_{\rm cn}$ and due to transition
to photon cooling stage.
The effect is illustrated in Fig.\ \ref{fig-cool}
which shows cooling curves (dots and dashes) for
two models of neutron SF (dots and dashes)
displayed in Fig.\ \ref{fig-sup} and parameterized
by Eqs.\ (\ref{Tc}) and (\ref{nN}).
In both cases $k_0=1$ fm$^{-1}$.
The dotted line in Fig.\ \ref{fig-sup} corresponds to 
the following set of parameters:
$T_0=2.915 \times 10^8$ K, 
$k_1=0.5664$ fm$^{-1}$, $k_2=2.767$ fm$^{-1}$,
$k_3=0.2965$ fm$^{-1}$.
Accordingly, $T_{\rm cn}$ has maximum 
$1.93 \times 10^8$ K
at $\rho=5.62 \times 10^{14}$ g cm$^{-3}$ and vanishes
at $\rho = 1.80 \times 10^{15}$ g cm$^{-3}$.
The dashed line corresponds to 
$T_0=6.461 \times 10^9$ K, $k_1=1.961$ fm$^{-1}$, $k_2=2.755$ fm$^{-1}$,
$k_3=1.30$ fm$^{-1}$.
Thus, $T_{{\rm c}n}$ has maximum 
$3.41 \times 10^8$ K
at $\rho=4.52 \times 10^{14}$ g cm$^{-3}$ and vanishes
at $\rho = 1.78 \times 10^{15}$ g cm$^{-3}$.

The dashed--curve SF
has somewhat larger maximum $T_{\rm cn}$ 
and appears at earlier cooling stage. 
As seen from Fig.\ \ref{fig-cool}, any mildly strong
neutron SF destroys our theoretical explanation
of observational data producing pronounced
breaks of the cooling curves. It is evident
that the weaker SF of neutrons violates
theoretical interpretation of older NSs.
It would be impossible to explain rather
high observed surface temperature of old PSR 1055--52
and RX J1856--3754
by our NS models with sufficiently strong SF of neutrons
(regardless of SF of protons).
The NS heat capacity would be strongly reduced,
neutrino emission due to Cooper pairing of neutrons
would be important, and
the NS would be much colder.
Our interpretation of all sources in Fig.\ \ref{fig-cool} would
not be violated only by a weak neutron SF
with maximum $T_{\rm cn} < 10^8$ K.
This result is in line with low values of $T_{\rm cn}$
obtained earlier (Yakovlev et al.\ \cite{yls99}) using 
simplified models of $T_{\rm cn}$ constant throughout the NS cores.
Note that Yakovlev et al.\ (\cite{yls99})
were unable to explain rather high values of $T_{\rm s}^\infty$
for PSR 1055--52. We can provide such an explanation
using NS models with somewhat different EOS and more
advanced physics input in the NS crust.

\section{Summary}
\label{sect-sum}

We have proposed an interpretation of
the observations using a simple model of cooling
NSs. 
The model
is consistent with observational data if the 
proton SF in dense matter is rather strong,
with the maximum of $T_{\rm cp} \ga 5 \times 10^9$ K
at $\rho \sim (2-3) \, \rho_0$ but decreases 
sharply at $\rho \ga \rho_{\rm D}$.
On the other hand,
the neutron SF must be weak, with the maximum
$T_{\rm cn} < 10^8$ K. Fixing the density profiles of
$T_{\rm cp}(\rho)$ and $T_{\rm cn}(\rho)$ we can explain
the observational data by cooling of NSs with the
same EOS and SF properties
but with different masses.
Our interpretation does not require any reheating
mechanism in cooling isolated NSs
(see, e.g., Yakovlev et al.\ \cite{yls99}, for references).

A strong SF of protons is required
to smooth transition from slow to fast cooling with
increasing NS mass. Otherwise it would be
difficult to explain rather low surface
temperatures of the Vela and Geminga pulsars.
A weakness of neutron SF
is necessary to avoid sharp falls of the NS surface
temperatures after onset of neutrino emission due to Cooper
pairing of neutrons.
Weak neutron SF is in favor of
a not too soft EOS in the NS core (softness would mean
strong attractive nn interaction and therefore strong
neutron pairing). Strong proton SF is
in favor of not too large symmetry energy
at high densities (too large symmetry energy would
mean large proton number density which would suppress
proton pairing). On the other hand, the
symmetry energy must be not too small to open
direct Urca process at $\rho > \rho_{\rm D}$.

The adopted density dependence $T_{\rm cp}(\rho)$
is not unique. We could choose different $T_{\rm cp}(\rho)$
profiles satisfying the above criteria
and interpret observations in the same manner
by cooling of NSs with somewhat different masses. 
Although we have used one particular EOS
in the NS core, about the same results would
evidently be obtained for a variety of EOSs.

In our model,
the colder NSs,
Vela and Geminga, are more massive ($M \sim 1.45 \, {\rm M}_\odot$),
while the hottest RX J0822--43 and warm and old
PSR 1055--52 and RX J1856--3754 are 
the least massive
($M \sim (1.1-1.3) \, {\rm M}_\odot$) of all NSs included in the analysis.
The mass range obtained is in good agreement
with the well--known range of masses of radio pulsars
in binary systems (Thorsett \& Chakrabarty \cite{tc99}).
Note that in our model
neutrons in the NS cores remain nonsuperfluid for $t \la 10^6$ yr. 
Furthermore, massive NSs,
$M \ga  1.5\,{\rm M}_\odot$, must undergo fast
cooling and be cold,
$T_{\rm s}^\infty \la 3 \times 10^5$ K
(the lowest solid curve in Fig.\ \ref{fig-cool}).
Thus, we could explain the existence of
cold middle--aged NSs, if such  objects  were discovered,
as massive NSs.
On the other hand, it would be easy
to modify the model to avoid prediction of these cold
NSs. This can be done by choosing
the EOS in the NS core
in which $\rho_{\rm D}$ is close to $\rho_{\rm c,max}$.
Another possibility is to adopt the $T_{\rm cp}(\rho)$ profile
with $T_{\rm cp}(\rho_{\rm c,max}) \ga (2-3) \times 10^9$ K.

Our model may seem oversimplified because
it neglects possible presence of other particles
(muons, hyperons, quarks) in the NS cores. 
However, despite its simplicity, it can explain observations 
of isolated NSs with only one adjustable parameter -- 
NS mass. 
We are planning to analyze more complicated models in future
publications. 

\begin{acknowledgements}
We are indebted to George Pavlov for very helpful discussions,
to Kseniya Levenfish for assistance at the
initial stage of this work and for helpful critical remarks,
and also to Fred Walter and to the referee, Slava Zavlin,
for useful critical remarks.
The work was supported partly by RFBR grant No.\ 99-02-18099 
and KBN grant No.\ 5P03D.020.20.
\end{acknowledgements}

\end{document}